# Multi-channel deep convolutional neural networks for multi-classifying thyroid disease

Xinyu Zhang[a], Vincent CS. Lee[a,∗], Jia Rong[a], James C. Lee[b,c], Jiangning Song[d], Feng Liu[e]

[a]*Department of Data Science and AI, Faculty of IT, Monash University, Clayton, Melbourne, VIC 3800, Australia*
[b]*Monash University Endocrine Surgery Unit, Alfred Hospital, Melbourne, VIC 3004, Australia*
[c]*Department of Surgery, Monash University, Melbourne, VIC 3168, Australia*
[d]*Biomedicine Discovery Institute and Department of Biochemistry and Molecular Biology, Monash University, Clayton, Melbourne, VIC 3800, Australia*
[e]*West China Hospital of Sichuan University, Chengdu City, Sichuan Province 332001, China*

## Abstract

Thyroid disease instances have been continuously increasing since the 1990s, and thyroid cancer has become the most rapidly rising disease among all the malignancies in recent years. Most existing studies focused on applying deep convolutional neural networks for detecting thyroid cancer. Despite their satisfactory performance on binary classification tasks, limited studies have explored multi-class classification of thyroid disease types; much less is known of the diagnosis of co-existence situation for different types of thyroid diseases. Therefore, this study proposed a novel multi-channel convolutional neural network (CNN) architecture to address the multi-class classification task of thyroid disease. The multi-channel CNN merits from computed tomography to drive a comprehensive diagnostic decision for the overall thyroid gland, emphasizing the disease co-existence circumstance. Moreover, this study also examined alternative strategies to enhance the diagnostic accuracy of CNN models through concatenation of different scales of feature maps. Benchmarking experiments demonstrate the improved performance of the proposed multi-channel CNN architecture compared with the standard single-channel CNN architecture. More specifically, the multi-channel CNN achieved an accuracy of $0.909 \pm 0.048$, precision of $0.944 \pm 0.062$, recall of $0.896 \pm 0.047$, specificity of $0.994 \pm 0.001$, and F1 of $0.917 \pm 0.057$, in contrast to the single-channel CNN, which obtained $0.902 \pm 0.004$, $0.892 \pm 0.005$, $0.909 \pm 0.002$, $0.993 \pm 0.001$, $0.898 \pm 0.003$, respectively. In addition, the proposed model was evaluated in different gender groups; it reached a diagnostic accuracy of 0.908 for the female group and 0.901 for the male group. Collectively, the results highlight that the proposed multi-channel CNN has excellent generalization and has the potential to be deployed to provide computational decision support in clinical settings.

*Keywords:*
Deep learning; Thyroid disease; Multi-channel CNN; Multi-class classification; Computer-aided diagnosis (CAD)

## 1. Introduction

In recent years, deep learning techniques have brought unprecedented opportunities and transformative breakthroughs in various fields. In clinical medicine, deep learning-based computer-aided diagnostic (CAD) systems have been shown to yield satisfactory, sometimes even superior, diagnostic accuracy and efficiency compared to experienced clinicians [1, 2, 3, 4]. Although the application of deep learning techniques on medical images is commonly used to enhance diagnostic performance, there is an ever-growing demand for more advanced deep neural networks to address more challenging scenarios. Currently, many existing studies highlight the competitive performance of deep convolutional neural networks

(CNN) when being applied to diagnose various diseases, such as breast cancer [5], kidney disease [6], and cardiac disease [7]. Those models produce decent results on binary classification tasks for detecting a particular disease; however, in many cases, they are inferior to human-level diagnosis, where expert clinicians usually incorporate more facets of domain knowledge to make a comprehensive diagnostic conclusion. Therefore, deep neural networks cannot be primarily approbated clinically due to the limitations of solely relying on binary classification tasks, the comprehensiveness of human-level diagnosis, and the interpretability of such "black-box" approaches.

With the aforementioned issues in mind, this paper highlights the potential capability of deep learning-based neural networks to approach human-level diagnosis to support clinicians making decisions more efficiently and comprehensively. A novel multi-channel CNN architecture was proposed in this study with those objectives. Specifically, the proposed multi-channel CNN will fuse feature maps generated through differ-ent convolutional kernels to produce an output; this upholds the increased diagnostic performance compared to the human-level diagnosis and also to the conventional CNN architectures. The proposed multi-channel deep CNN was inspired by the characteristics of computed tomography (CT). CT is an imaging modality that is widely available and commonly used to visualize solid organs and their associated diseases [8, 9]. The acquisition of CT images is less operator-dependent and more protocolized than some other diagnostic imaging modalities, such as ultrasonography. This property makes CT images rather appropriate for CAD implementations. This paper focuses on the most common endocrine organ, the thyroid gland [10], to evaluate the proposed model. The results reached the current state-of-the-art performance for thyroid disease diagnosis. In addition, the proposed multi-channel architectures were compared to the conventional single-channel CNN. The multi-channel architecture was also evaluated on gender disparity to examine the model's generalizability. Moreover, it is also feasible for the model to be generalized to diagnose different diseases.

∗Corresponding author at: Machine Learning & Deep Learning Discipline Group, Department of Data Science & Artificial Intelligence, Faculty of Information Technology, Monash University, Clayton Campus, 20 Exhibition Walk, Woodside (T & D) Building, VIC 3800, Australia; office 2.63, Tel: +61399052360, Email: vincent.cs.lee@monash.edu

Thyroid diseases can be broadly classified into functional and neoplastic [11]. Functional thyroid diseases include Graves' disease, lymphocytic (Hashimoto's) thyroiditis, which can lead to hyperthyroidism or hypothyroidism. Neoplastic thyroid conditions, such as cysts, adenomas, and multi-nodular goiter, can be further categorized into benign and malignant [11]. Neoplastic thyroid disease is widespread worldwide, for which more than 50% of adults have thyroid nodules [12]. Additionally, functional and neoplastic diseases can co-exist, for example, toxic multi-nodular goiter, toxic adenoma, and malignancy in the setting of thyroiditis; so can different neoplastic diseases, leading to a multi-nodular goiter. In this regard, existing studies typically focus on binary classification tasks, including classifying between hypothyroidism and hyperthyroidism [13, 14]. Additionally, other studies attempted to use medical images for distinguishing cancerous thyroid nodules from benign nodules [15, 16, 17]. There is a scarcity of studies on multi-classifying neoplastic thyroid diseases or outlining the circumstance where an individual patient might suffer from various types of thyroid diseases simultaneously in the context of deep learning.

In the clinic, neoplastic thyroid disease is commonly diagnosed through rigorous procedures, involving blood tests, ultrasonography, CT, fine-needle aspiration cytology (FNAC), and excisional biopsy if necessary. Although FNAC is currently regarded as the gold standard for a pre-operative diagnostic test for malignancy, 20 - 30% are reported as nondiagnostic or indeterminate due to its inherent limitations [18]. Patients without a definitive diagnosis on FNAC preoperatively are more likely to receive sub-optimal initial surgery [19]. The process of reaching a diagnosis for a clinically significant thyroid nodule is complex, nuanced, and costly. From the patient's perspective, there is room for improvement to minimize cost and anxiety. Therefore, this paper proposed the multi-channel CNN to help streamline the diagnostic process for thyroid disease, aiming for a highly efficient and accurate CAD system. We believe this system has the potential to be integrated into the clinical workflow to guide primary care physicians in deciding if the specialist referral is warranted.

This paper facilitates our understanding of deep learning applications in the medical domain, in which we witness a growing interest in such explorations. Overall, this paper makes the following contributions:

· To our knowledge, this study is the first to apply multi-channel CNN in the diagnosis of thyroid disease types. The multi-channel CNN brought the current state-of-the-art performance for thyroid disease diagnosis. This CAD system more closely emulates the human-level diagnostic process by going beyond a binary classification system, thus taking a step towards seamless integration into clinical practice.

· Another novel finding of this study is the comparable accuracy of the multi-class CAD system to that of experienced clinicians. This important advance bridges the gap left by earlier studies with binary outcomes. Additionally, this paper offers patients the most decent and comprehensive diagnostic results, emphasizing the application of CNN in detecting disease co-existence possibility.

· This study evaluates the impact of kernel size selection on CNN performance. The proposed multi-channel model was evaluated through different kernel size selections, and it was also compared to the single-channel CNN. This study also recommends the most stable kernel size combination for the CNN model.

· This study assesses the CNN performance on gender disparity. The proposed CNN was evaluated with respect to different gender groups, and the model demonstrates promising generalizability, suggesting its potential to assist in diagnosing gender-specific thyroid diseases.

· This paper highlights the value and importance of incorporating patient-centered design into deep learning-based CAD applications. The proposed model fits individual diagnoses, thereby making the diagnostic decisions tailored. If proven useful clinically with accuracy confirmed in larger, prospective studies, the CAD system may streamline the diagnostic process for some patients by avoiding biopsy or diagnostic surgery. It also has the potential to increase the efficiency of a clinician's workflow by suggesting preliminary diagnoses.

· This work is reproducible in the sense that all the related protocols and partial datasets are publicly available through GitHub.

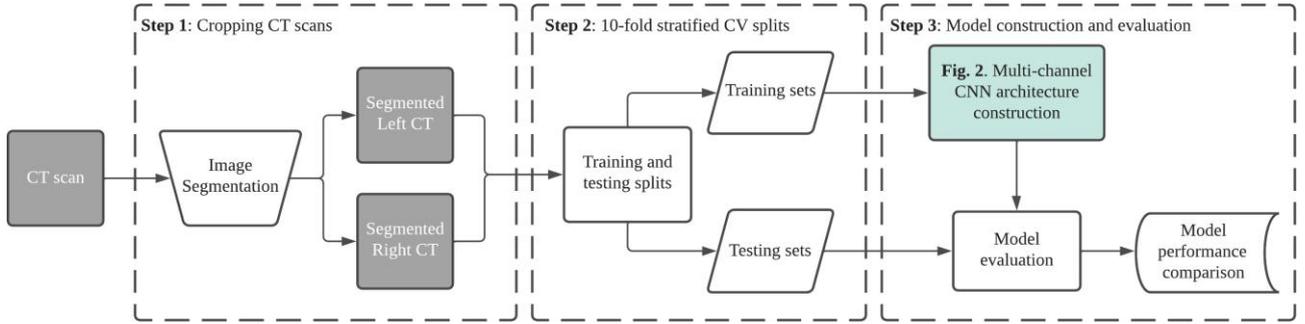

Figure 1: Multi-channel CNN construction framework for thyroid disease diagnosis

## 2. Related work

This section categorizes the existing works into multi-channel CNN-related literature studies and thyroid disease classification-relevant studies.

### 2.1. Multi-channel CNN-related studies

Conventional CNN architectures like VGG models were prestigious in the last few decades due to their outstanding performance on natural image classifications [20]; however, researchers nowadays accentuate that those models are not advantageous enough to effectively handle medical images which share similar but less complex structures compared to natural images [21]. In order to emphasize the characteristics and excavate useful features from medical images, researchers tend to shift their focus to building CNN models that can bring captivating classification performance on medical images. Among various types of advanced CNN architectures, multi-channel CNN highlights the significant impact between image characteristics and CNN performance [22], which is the most rational design. Multi-channel CNN architecture has been commonly used in other research fields, such as for sentiment analysis [23] and hyperspectral data classification [24].

In the medical field, multi-channel CNN has been applied on several occasions. For instance, [25] adopted multiple instance learning with multi-channel CNN to identify infants who might have the risk of autism disease. In their work, the magnetic resonance image was split into different patches containing suspicious features, which were subsequently input into the multi-channel CNN with the same kernel size to perform classification. As a result, their method reached an accuracy of 0.724. A similar approach was conducted by [26] for brain disease detection. A consensus of several existing studies is that multi-channel CNN architectures can enhance diagnostic performance for a particular disease [27, 22].

In addition, a recent study evaluating different kernel sizes for histopathology slides reported a moderately strong association between kernel size and CNN performance [28]. Similarly, [29] showed that multi-channel CNN architectures could increase diagnostic performance, and the kernel size choice for multi-channel CNN architectures might be relevant for the binary classification performance. However, the performance of multi-channel CNN on multi-classification tasks is yet to be explored. Hence, this study proposes a novel multi-channel CNN architecture that renovates kernel size selection adopts CT scans to diagnose multi-class thyroid diseases in a patient-specific manner.

### 2.2. Thyroid disease classification studies

As the largest endocrine gland in the human body, the thyroid plays an essential role in regulating our daily metabolism [30]. With the continuously increased cancer incidence rates, awareness of thyroid diseases is gradually increasing among the clinicians, patients, and the public [31, 32]. In recent years, the binary classification task using deep learning techniques via ultrasonography for thyroid cancer detection has dominated the literature in this field.

[17] proposed a fused CNN to distinguish between benign and malignant thyroid nodules via ultrasound images and reached an accuracy of 83.02%. [33] applied ResNet18 to 4, 509 ultrasound images, and reached a diagnostic accuracy of 0.8388. [34] used 589 ultrasound nodules for classifying between benign and malignant classes. Their results show that they obtained the area under the curve values of 0.845, 0.835, and 0.850, respectively, for three different CNN models. There also exist many other studies on thyroid nodule classifications based on ultrasound images using CNN, all of which reached relatively satisfactory diagnostic accuracy [35, 36, 37, 38, 39, 40].

Many other studies adopted different medical imaging modalities for thyroid cancer detection. For instance, [41] used hyperspectral images from 44 patients to detect papillary thyroid carcinoma, and accordingly, they reached an area under curve of 0.85. In another work, [42] utilized 995 CT images on ResNet50 and achieved a testing accuracy of 0.904. [3] compared the CNN performance with two radiologists on 986 CT images, and their results showed the CNN models outperformed the radiologists. It is important to note that this study represents the first attempt to make use of CT characteristics in designing CNN architectures. Our proposed model selects CT scans because it provides the integral appearance of the thyroid gland; meanwhile, CT scans may also help determine the disease co-existence diagnostic decisions.

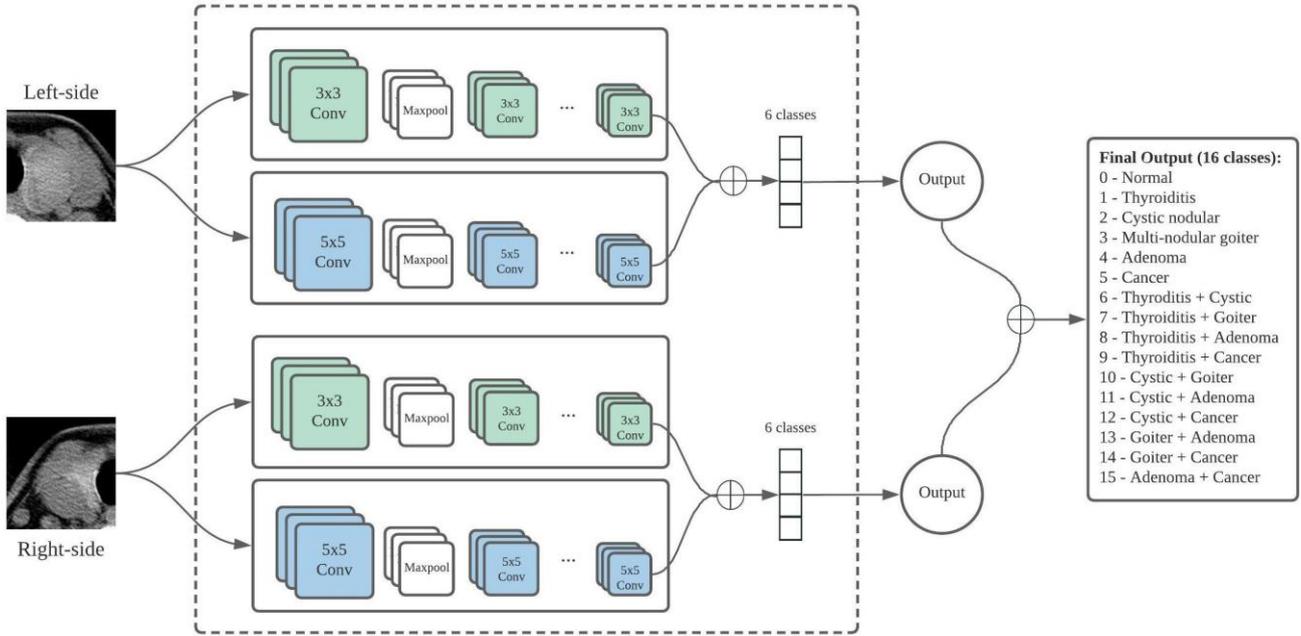

Figure 2: Multi-channel multi-class classification CNN architecture

Nevertheless, among all the existing studies, no piece of literature contemplates and incorporates the circumstance in which various types of thyroid diseases can co-exist; that is, the situation where sometimes a patient might have one side of the thyroid gland appears to have benign nodules, while the other side of the gland appearing to have cancerous nodules. Another issue is that the multi-classification task of incorporating all thyroid diseases is lacking in the existing studies. Therefore, this research bridges the gaps and proposes a novel computational framework for the multi-class classification of thyroid diseases using multi-channel CNN architecture. The proposed multi-channel multi-class CNN allows fusion of the diagnoses made by different thyroid lobes, thereby addressing the challenging issue of detection disease co-existence while providing a human-level diagnostic decision.

## 3. Methodology

With the objective to diagnose thyroid disease through deep learning techniques, this section elaborates the proposed flowchart and explains the designed multi-channel CNN architecture.

### 3.1. Multi-channel multi-class CNN framework

The overall workflow of the proposed multi-channel CNN is presented in Fig. 1. Here, we adopted CT scans to develop a deep learning-based approach rather than ultrasound images. Existing studies have mainly used ultrasound images for thyroid disease diagnosis coupled with deep learning-driven algorithms [37, 38, 39]. Due to the internal details and high resolution provided by ultrasonography to the human eye, it is the preferred modality for the characterization of thyroid nodules in the clinical setting. However, ultrasound images usually focus on a single nodule at a time, and the quality of the images is very much operator-dependent. On the other hand, protocolized acquisition of CT images can provide images with a consistent quality, as well as an overview of the entire thyroid gland and its surrounding structures. Therefore, we have chosen to use CT images as the input material in this study.

Since each of the two thyroid lobes can have a different diagnosis, segmenting the overall CT scan is necessary. By segmenting the CT images into left and right lobes, a different label can be applied to each. Segmentation also allows the CAD system to diagnose separately to each lobe. With this increased complexity, the overall diagnostic ability of the system can mimic real-life diagnoses made by a clinician more closely.

The first step of the flowchart segments the acquired CT scans into the left and right lobes. A dominant class label will be assigned to each image with the cropped images. The second step was to apply the 10-fold stratified cross-validation (CV) for training and testing splits with all the segmented CT images. In the third step, the segmented left and right-side CT scans from the training sets will be fed into the constructed multi-channel CNN architecture, while the testing sets will be used to evaluate the proposed CNN architecture. The model will be evaluated through seven major measures, including accuracy, precision (also known as positive predictive value - PPV), recall (also known as sensitivity), specificity, negative predictive value (NPV), F1, and variance. Finally, different combinations of the kernel sizes for constructing the multi-channel will be compared in terms of these performance metrics.

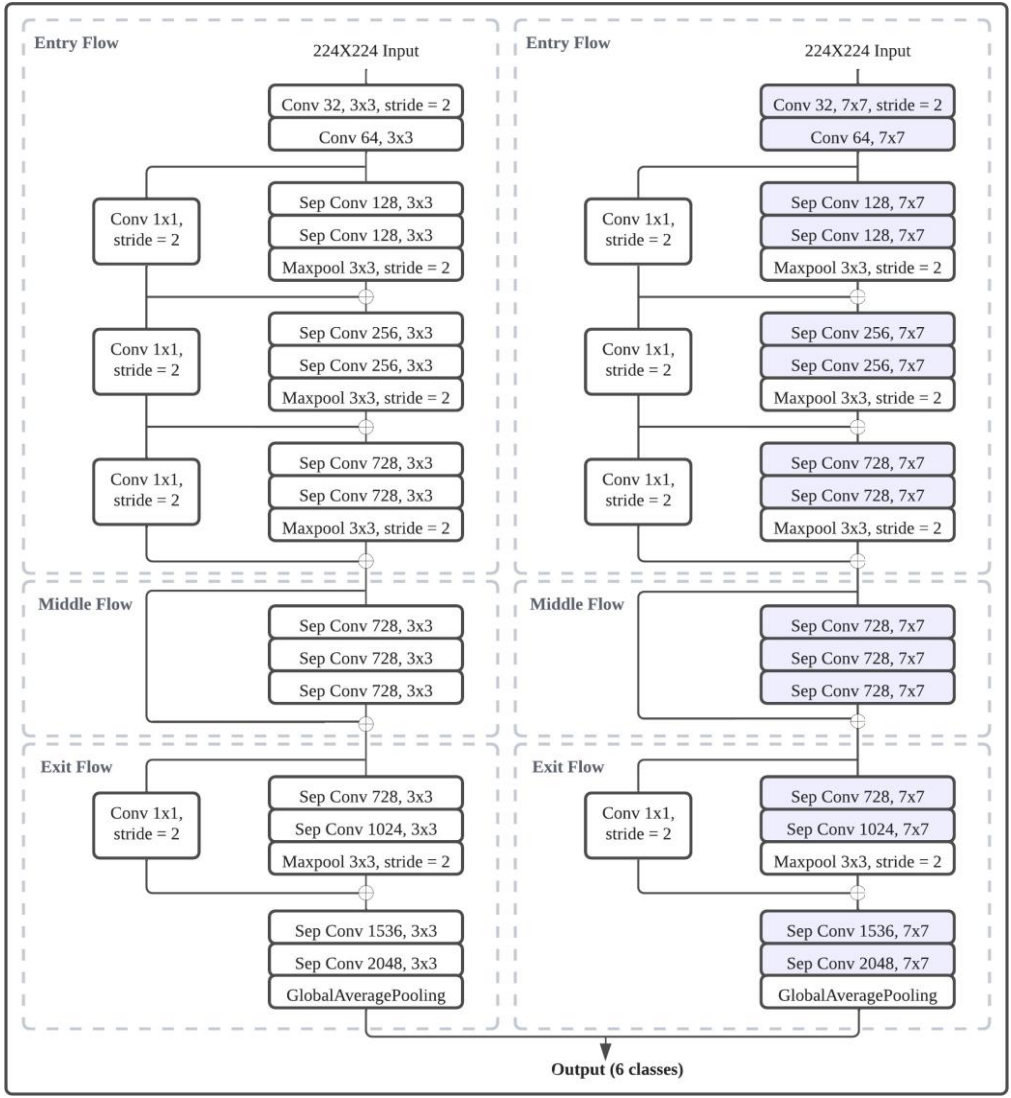

Figure 3: Dual-channel Xception architecture (adapted from [43])

The detailed multi-channel multi-class CNN architecture is illustrated in Fig. 2. The key to constructing a successful multi-channel CNN is the choice of appropriate kernel sizes for the convolutional operations. Kernel size is remarkably influential to a CNN performance [44, 28]. Kernel size convolves around feature maps and determines the size of the receptive fields. A larger kernel choice will prompt having more abstract features to be extracted from the input image, while a smaller kernel size will learn more detailed textures. Deriving from this, we proposed a multi-channel CNN, which merges the feature maps generated through a smaller choice of kernel size with a larger kernel size to obtain an intermediate feature map. In doing so, more informative features can be learned from the input image in a balanced manner.

The most commonly used kernel sizes for the CNN architectures are $1 \times 1$, $3 \times 3$, $5 \times 5$, $7 \times 7$, and $9 \times 9$ [45, 28]. [28] revealed by the evaluation of the impact of kernel size selection on the CNN performance via histopathological images, and the results indicated that the CNN would have the highest validation errors when applying the kernel size of $9 \times 9$; thus, the $9 \times 9$ kernel size was excluded in this research as it cannot learn sufficient features for making predictions.

Compared to using all four kernel sizes in different chan-nels, the dual-channel architecture is more appropriate as it requires less computational resources to train; besides, dual-channel is adequate for handling medical images as they have simpler structures. As such, we implemented the dual-channel architecture to include one smaller kernel and the other larger kernel for classification tasks, with the purpose of improving both the diagnostic accuracy and efficiency. In this study, $1 \times 1$ and $3 \times 3$ kernels were considered as 'smaller kernel sizes', while $5 \times 5$ and $7 \times 7$ are considered as 'larger kernel sizes'. We further performed ablation studies to evaluate the impact of different combinations of the kernel sizes, including $1 \times 1$ with $5 \times 5$, $1 \times 1$ with $7 \times 7$, $3 \times 3$ with $5 \times 5$, and $3 \times 3$ with $7 \times 7$. Additionally, we also compared the performance the dual-channel architectures to that of the single-channel architecture.

Importantly, the inputs of the multi-channel architectures for multi-classification to detect thyroid disease included the segmented left and right thyroid gland CT scans from a particular patient. Each of the left and right sides of the CT scans will be processed by the dual-channel with different kernel sizes. Subsequently, the overall CNN will further fuse the processed features into an intermediate feature map and generate a classification vector that includes six classes of thyroid disease (i.e., normal, thyroiditis, cystic nodule, multi-nodular goiter, adenoma, and cancer). Although thyroid disease can be classified into more than those classes, this study incorporates the six most commonly seen types for evaluations. Then, these generated vectors from both left and right sides were further concatenated into a $16 \times 16$ matrix, indicating the overall status of the gland. The 16 classes include the aforementioned six diseases (if one side or both lobes had the disease), as well as ten different combinations of the thyroid diseases of different lobes (see Fig. 2). Accordingly, the final output would be the class label for a given CT scan for a specific patient, specifying the disease type of the whole thyroid gland. This architecture allows the diagnosis of individual patients each time in a patient-specific manner.

### 3.2. CNN architecture – Xception

With the emergence of deep learning techniques, CNN has brought remarkable performance in computer vision tasks [46]. More accurate and efficient CNN architectures are highly desirable to better handle the complexity of the classification and object detection tasks. The classical CNN models (i.e., AlexNet, VGG) require the stacking of more layers to increase its performance; this, in turn, requires high-performance computational resources necessary for training the model and sometimes might face gradient explosion concerns [47]. Therefore, more advanced architectures have been designed by researchers to solve gradient explosion issues whilst preserving the model complexity to solve more challenging tasks with increased accuracy and efficiency, such as ResNet [47], Inception [45], and Xception [43].

Xception was proven to outperform VGG models, ResNets, and Inception models with faster training speed on ImageNet [43]; this makes the model well-suited for the medical fields as medical data are of massive volumes and heterogeneous complexity. Training such data with long-established CNN models (i.e., AlexNet or VGG) is quite time-consuming, so as to increase the efficiency, Xception was chosen in this study as the base model for building the multi-channel CNN (see Fig. 3).

Furthermore, Xception also achieved the most accurate classification performance among the aforementioned architectures. In the medical domain, diagnostic accuracy is another major factor that needs to be taken into consideration for CAD design in addition to the efficiency. Therefore, Xception was selected in this study due to its superior performance in identifying spatial correlations of abnormal lesions among other structures in a given image [43]. Overall, the characteristics of Xception make it ideal for processing a large amount of medical data with enhanced accuracy and efficiency.

## 4. Experiments

The proposed multi-channel CNN architecture was evaluated through CT images, and this section introduces the acquired data set and explains the experimental parameters setting.

### 4.1. Data set acquisition

With the ethics approved by the Monash Human Ethics Committee, we obtained consent from a tertiary hospital in Sichuan Province, China. In order to preserve the anonymity of the patients, the patient records have been de-identified for this study.

This study retrospectively recruited consecutive patients undergoing thyroid disease-related workup from August 2019 to August 2020. As a result, a total of 576 patients were included in this study. Table 1 displays the distribution of the patients' demographical information. As can be seen, a large group of patients was from the age group 55 to 75 with a percentage of 42.54%. Moreover, a substantial portion of the patients were females, with a percentage of 76.22%. Additionally, more than 80% of the pathological results were benign, including thyroiditis, cystic nodule, multi-nodular goiter, and adenoma.

From 576 patients, a total of 5$mm$ sliced 977 contrast CT images were segmented into left and right lobes. Fig. 4 shows an example of the segmented left and right-side thyroid CT images in different classes. Definitive diagnoses of each lobe based on postoperative histopathology were used to label all the CT images; images were excluded from this study if no histopathological results were available. In addition, the dominant class was assigned based on the severity of the disease, which we consider cancer is the most severe type, followed by adenomas, goiters, cystic, thyroiditis, and normal. The detailed CT scan distribution for each of the six classes is presented in Table 2.

Table 1: Description of the dataset used in this study

| Demographics | |
|---|---|
| **Age** | **Percentage (%)** |
| Below 18 | 0.35 |
| 18 - 35 | 10.24 |
| 35 − 55 | 40.45 |
| 55 − 75 | 42.54 |
| 75 + | 6.42 |
| **Gender** | |
| Male | 23.78 |
| Female | 76.22 |
| **Pathology** | |
| Benign | 81.78 |
| Malignant | 18.22 |

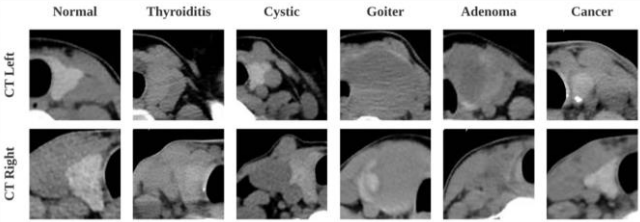

Figure 4: Sample left and right-side CT scan for the six classes

### 4.2. Experimental setting

After segmenting the CT images into left and right lobes through a Python-based CT image segmentation generator (also available through GitHub), we labeled and resized all the images to $224 \times 224$ for the sake of consistency. Set the input image set as $X \in \mathbb{R}^{224 \times 224}$, label annotated as $y \in \{0, 1, 2, 3, 4, 5\}$; specifically, 0 denotes the "normal" class, 1 denotes "thyroiditis", 2 denotes "cystic nodule", 3 denotes "multi-nodular goiter", 4 denotes "adenoma", and 5 denotes "cancer", respectively. The remaining ten classes indicate the combinations of the diseases, which can be found in Fig. 5.

In order to highlight the disease co-existence situation, our labeling process was rigorous. Specifically, if a patient has normal thyroid lobes for both left and right sides, then the patient is considered "normal"; similarly, if the patient has the same disease for both sides, the corresponding images will be labeled with the dominating disease class. Moreover, if the patient has one side as "normal", while the other side has other types of thyroid disease, then the patient will be classified as the dominating disease class. For example, if the patient has a left-side CT image that appears to have a normal lobe, while the right-side has cancerous nodules, then the overall diagnosis for this patient would be "cancer". In addition, the different combinations of the thyroid disease were given based on the dominant class for each side of the CT image. Thus, we have a total of 16 classes to be classified; apart from the primary six classes, there are thyroiditis with cystic nodule, thyroiditis with goiter, thyroiditis with adenoma, thyroiditis with cancer, cystic with goiter, cystic with adenoma, cystic with cancer, goiter with adenoma, goiter with cancer, and adenoma with cancer (Fig. 2).

### 4.3. Data imbalance

Standard image augmentation techniques such as cropping, zooming, flipping, and rotating are not appropriate in this case. Therefore, data augmentation techniques were not applied in this study since the model requires rigorous amounts of the input of CT images for both left and right sides, where the input volumes for multi-channels must be the same. Nevertheless, in order to address the class imbalance issue, we have incorporated the stratified CV technique to generate unbiased classification results [48]. The stratified CV approach splits each stratified fold so that each fold would retain approximately the same ratio of class labels as the observations; this allows the variance among all the predictions to be reduced, making the average error estimate representable and reliable [49]. Thus, the stratified CV technique is selected in all the experiments rather than the standard CV technique.

In addition, we applied the categorical cross-entropy (CCE) as the loss function. The CCE is usually applied in multi-class classification tasks as it can be weighted based on different classes. CCE can adapt the penalty of a probabilistic false-negative rate for a given class [50], making it appropriate for this multi-classification study. The CCE is calculated using Eq. 1. With the one-hot encoded labels, the last fully connected layer will produce a vector indicating the possibility of each class label, and the $s_p$, in this case, denotes the predicted score for the specific class, $s_j$ is the inferred score for each class in $C$, while $C$ is the total number of classes.

$$CCE = -log(\frac{e^{s_p}}{\sum_j^C e^{s_j}}) \qquad (1)$$

Moreover, the predicted score for each class is calculated through Eq. 2, where $Y_k$ is the predicted probability for $k_th$ image, $f$ is the feature map size (see Eq. 3), $w_k$ is the weight for the $k_th$ feature map through the corresponding convolutional operation, and $X$ is the encoded input image.

$$Y_k = f(w_k \times X) \qquad (2)$$

In addition, Eq. 3 demonstrates the feature map size where $n^h$, $n^w$, and $n^c$ are the height, weight, and channel of the input image, $f$ is the kernel size, and $s$ is the stride size for the convolutional operations.

$$f = (\frac{n^h - f}{s} + 1) \times (\frac{n^w - f}{s} + 1) \times n^c \qquad (3)$$

### 4.4. Parameters setting and evaluation metrics

All the experiments applied the Adam optimizer, with an initial learning rate of $1 \times 10^{-2}$, then the learning rate was gradually updated during the fine-tuning procedure. The CNN model was found to perform most stably with a learning rate of $1 \times 10^{-5}$, and it was set as fixed afterwards. During each training iteration, the batch size was set to 2. In order to evaluate the proposed multi-channel CNN architectures, four performance measurements were utilized in this study, and they are calculated through Eq. 4 to 9 (TP, True Positive; TN, True Negative; FP, False Positive; FN, False Negative; k is number of folds).

Table 2: Distribution of CT scans in the six classes of thyroid diseases

| Side | Class | | | | | | Total |
|------|--------|-------------|--------|--------|---------|--------|-------|
|      | Normal | Thyroiditis | Cystic | Goiter | Adenoma | Cancer |       |
| Left | 199 | 68 | 299 | 178 | 55 | 178 | 977 |
| Right | 217 | 72 | 308 | 179 | 47 | 154 | 977 |

$$Accuracy = \frac{1}{k} \sum_{i=1}^{k} \frac{TP_i + TN_i}{TP_i + TN_i + FP_i + FN_i} \quad (4)$$

$$PPV = \frac{1}{k} \sum_{i=1}^{k} \frac{TP_i}{TP_i + FP_i} \quad (5)$$

$$Sensitivity = \frac{1}{k} \sum_{i=1}^{k} \frac{TP_i}{TP_i + FN_i} \quad (6)$$

$$Specificity = \frac{1}{k} \sum_{i=1}^{k} \frac{TN_i}{TN_i + FP_i} \quad (7)$$

$$NPV = \frac{1}{k} \sum_{i=1}^{k} \frac{TN_i}{TN_i + FN_i} \quad (8)$$

$$F1 = 2 \times \frac{Precision \times Recall}{Precision + Recall} \quad (9)$$

In addition, all the experiments were performed under the same computational environment, involving a 64-bit Windows 10 Pro desktop, which had an Intel Core i7-9700 processor with 16 gigabytes of memory and a GeForce GTX 1050 GPU.

## 5. Results

This section examined four different kernel size combinations for constructing multi-channel CNN architectures.

Table 3 provides the performance of the four types of multi-channel multi-class CNN models in terms of their accuracy, precision (PPV), recall (sensitivity), specificity, NPV, and F1 scores. When $1 \times 1$ and $5 \times 5$ convolutions were used in the multi-channel architecture, the model reached a classification accuracy of 0.905 with a variance of 0.048 on the 10-fold stratified CV. Specifically, the precision, recall, specificity, NPV, and F1 scores for the $1 \times 1$ and $5 \times 5$ combinations were 0.925, 0.889, 0.993, 0.993, and 0.904, respectively. When the $1 \times 1$ and $7 \times 7$ convolutions were applied, the highest accuracy among the four combinations was obtained, which was 0.909; however, the recall score was relatively lower than that of the other combinations, which was 0.896. Additionally, with the kernel choices of $3 \times 3$ and $5 \times 5$, the model achieved an accuracy of 0.906, which was slightly lower than that of $1 \times 1$ and $7 \times 7$. For the combination of the kernel sizes of the $3 \times 3$ and $7 \times 7$, the model had the lowest accuracy, precision, specificity, NPV, and F1 among all four choices, which were 0.9, 0.907, 0.992, 0.992, and 0.903, respectively. Nevertheless, the highest recall score of 0.907 with a variance of 0.06 was achieved with $3 \times 3$ and $7 \times 7$ kernel combination. Since the performance of all the models was benchmarked by conducting the 10-fold stratified CV, the variance for the four performance metrics was also calculated and provided in Table 3.

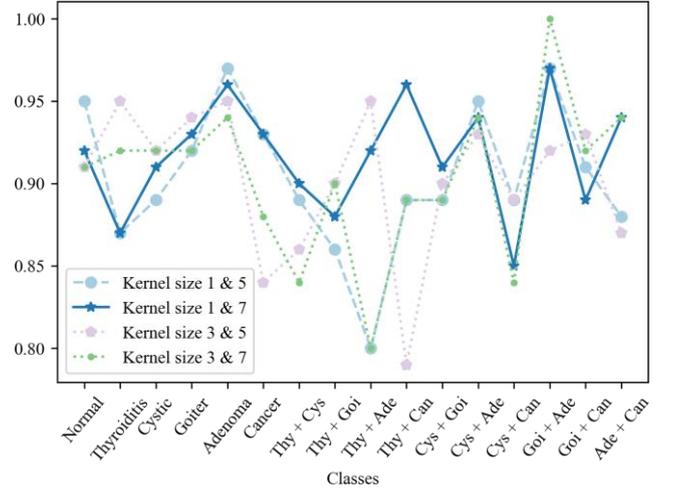

Figure 5: Mean F1 scores for multi-channel multi-class CNN architectures on 10-fold CV test

In addition, the mean F1 scores for the 16 classes on the 10-fold stratified CV are shown in Fig. 5. As can be seen, the most stably performing model achieved the highest accuracy when kernel sizes $1 \times 1$ and $7 \times 7$ convolutions were applied. Specifically, in the $1 \times 1$ and $7 \times 7$ multi-channel models reached mean F1 scores of higher than 0.9 for classifying 12 out of 16 classes, which include normal, cystic, goiter, adenoma, cancer, thyroiditis with cystic, thyroiditis with adenoma, thyroiditis with cancer, cystic with goiter, cystic with adenoma, goiter with adenoma, and adenoma with cancer. In contrast, the least F1 score of 0.85 was obtained for classifying the cystic and cancer class. Besides, the lowest mean F1 scores of the four models were also presented for the thyroiditis with adenoma class. The combination of the kernel sizes $3 \times 3$ and $7 \times 7$

Table 3: Multi-class classification performance of the multi-class CNN model

| Kernel size | Accuracy | Precision (PPV) | Recall (Sensitivity) | Specificity | NPV | F1 |
|---|---|---|---|---|---|---|
| 1 & 5 | 0.905 ± 0.048 | 0.925 ± 0.049 | 0.889 ± 0.047 | 0.993 ± 0.001 | 0.993 ± 0.003 | 0.904 ± 0.052 |
| 1 & 7 | **0.909 ± 0.048** | **0.944 ± 0.062** | 0.896 ± 0.047 | **0.994 ± 0.001** | **0.994 ± 0.002** | **0.917 ± 0.057** |
| 3 & 5 | 0.906 ± 0.053 | 0.918 ± 0.052 | 0.894 ± 0.053 | 0.993 ± 0.005 | 0.993 ± 0.004 | 0.904 ± 0.055 |
| 3 & 7 | 0.900 ± 0.059 | 0.907 ± 0.060 | **0.907 ± 0.060** | 0.992 ± 0.007 | 0.992 ± 0.004 | 0.903 ± 0.061 |

appeared to be the most fluctuating model (Fig. 5).

## 6. Ablation studies

In order to evaluate the performance of the proposed multi-channel CNN architectures, we conducted three ablation studies in this study, including baseline model selection, multi-channel and single-channel CNN comparison, and gender disparity comparison.

### 6.1. Baseline model selection

Xception, as the baseline model, was compared to InceptionV3 and DenseNet121, which are the two commonly adopted CNN models for thyroid disease diagnosis [42, 51, 52]. The three CNN architectures were applied to the multi-class classification task for thyroid disease diagnosis.

It should be noticed here that the disease co-existence situation was not included in this stage, where the basic six classes were evaluated relative to different lobes. Specifically, Xception reached the highest precision, recall, and F1 scores for multi-classifying thyroid CT both in the left and right lobes. Table 4 shows that Xception reached a precision of 0.974 and 0.955 in precision for both lobes, recall of 0.947 and 0.927, F1 scores of 0.948 and 0.940, respectively. Performance comparison with the baseline model proves that adopting Xception for developing a multi-channel architecture is the most appropriate choice.

### 6.2. Performance comparison between single-channel and multi-channel CNNs

The multi-channel architectures were compared to the single-channel architectures (i.e., with the kernel size as of $3 \times 3$, $5 \times 5$, and $7 \times 7$ in Table 5). In this stage, the disease co-existence circumstance was taken into consideration; therefore, the classification results from the left and right lobes were fused.

Fig. 6 illustrates the performance comparison of the multi-channel (i.e., combination of the kernel size of $1 \times 1$ and $7 \times 7$) and single-channel architecture (i.e., $7 \times 7$ kernel). The results show that the multi-channel CNN architecture achieved increased accuracy, precision, specificity, and F1 scores when compared to the single-channel architectures. Although the sensitivity was relatively lower for the multi-channel architecture, its overall performance is superior to the single-channel architecture, suggesting the feasibility of applying the multi-channel architecture in thyroid disease diagnosis.

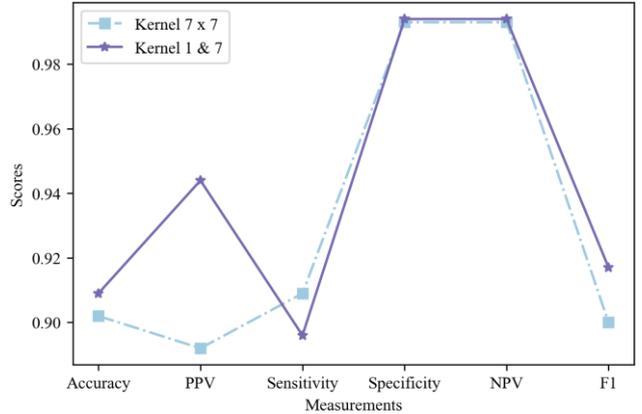

Figure 6: Averaged six performance metrics comparison of multi-channel and single-channel CNN models

### 6.3. Gender disparity comparison

In this section, we further evaluated the performance of the proposed multi-channel CNN architecture with respect to female and male groups. Specifically, the best-performing architecture (i.e., $1 \times 1$ with $7 \times 7$ combination) was evaluated. The numbers of the input images for the female and male groups were 774 and 203, respectively. Table 6 displays the results. For both gender groups, the classification accuracy of the multi-channel CNN architecture is promising. Specifically, for the female group, the multi-channel architecture reached 0.908, 0.931, 0.898, 0.994, 0.994, and 0.912 for accuracy, precision, recall, specificity, NPV, and F1, respectively. The male group had the corresponding scores of 0.901, 0.954, 0.9, 0.992, 0.992, and 0.913, respectively.

Additionally, Fig. 7 demonstrates the averaged F1 scores for female and male groups in the 16 thyroid disease classes on the 10-fold stratified CV. It should be noticed that the classes "goiter with adenoma" and "adenoma with cancer" were absent for the male groups; thus, no results were available for these two classes.

Fig. 7 also exhibits that the multi-channel CNN architecture could generalize well to different gender groups with an F1 score of larger than 0.8 for almost all the classes. Although the input image scale for the male groups was much smaller, the model also achieved an outstanding performance. Therefore, the proposed multi-channel CNN model has the potential to be further extended to different diseases due to its excellent

Table 4: Baseline model comparison for multi-class classification of thyroid disease

|  |  | **Results** | | |
|  |  | InceptionV3 | DenseNet121 | Xception |
|---|---|---|---|---|
| **CT Left** | Precision | 0.668 ± 0.005 | 0.813 ± 0.002 | **0.947 ± 0.001** |
|  | Recall | 0.735 ± 0.008 | 0.790 ± 0.012 | **0.947 ± 0.001** |
|  | F1 | 0.698 ± 0.003 | 0.795 ± 0.002 | **0.948 ± 0.001** |
| **CT Right** | Precision | 0.628 ± 0.003 | 0.768 ± 0.002 | **0.955 ± 0.002** |
|  | Recall | 0.710 ± 0.023 | 0.795 ± 0.012 | **0.927 ± 0.001** |
|  | F1 | 0.665 ± 0.001 | 0.778 ± 0.005 | **0.940 ± 0.001** |

Table 5: Single-channel CNN performance

| Kernel | Accuracy | Precision (PPV) | Recall (Sensitivity) | Specificity | NPV | F1 |
|---|---|---|---|---|---|---|
| $3 \times 3$ | 0.880 ± 0.004 | 0.904 ± 0.005 | 0.875 ± 0.004 | 0.991 ± 0.008 | 0.992 ± 0.005 | 0.888 ± 0.004 |
| $5 \times 5$ | 0.900 ± 0.001 | **0.905 ± 0.005** | 0.894 ± 0.003 | 0.992 ± 0.007 | **0.993 ± 0.003** | 0.899 ± 0.003 |
| $7 \times 7$ | **0.902 ± 0.004** | 0.892 ± 0.005 | **0.909 ± 0.002** | **0.993 ± 0.001** | 0.993 ± 0.005 | **0.900 ± 0.001** |

Table 6: Gender disparity comparison

| Gender | Accuracy | Precision (PPV) | Recall (Sensitivity) | Specificity | NPV | F1 |
|---|---|---|---|---|---|---|
| Female | 0.908 ± 0.011 | 0.931 ± 0.003 | 0.898 ± 0.006 | 0.994 ± 0.007 | 0.994 ± 0.004 | 0.912 ± 0.003 |
| Male | 0.901 ± 0.005 | 0.954 ± 0.015 | 0.900 ± 0.017 | 0.992 ± 0.001 | 0.992 ± 0.001 | 0.913 ± 0.010 |

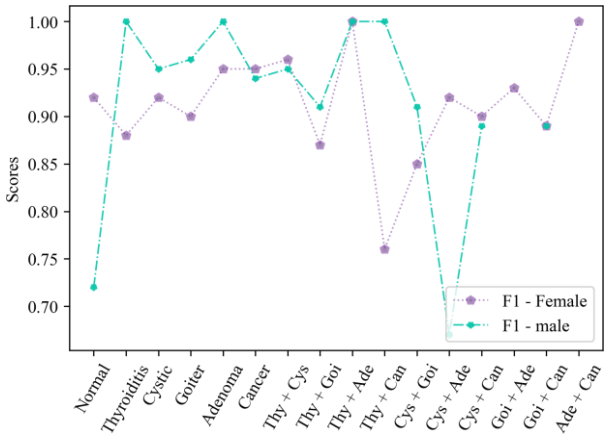

Figure 7: Performance comparison of the 16 thyroid disease classes in terms of the average F1 scores with respect to gender disparity on 10-fold CV

generalizability.

## 7. Discussion

Deep learning-based techniques have been increasingly utilized in the clinical domain to detect various diseases. In the case of thyroid disease diagnosis, deep CNN models were typically used to distinguish between benign and malignant nodules [4, 3]. Nevertheless, existing studies focus on classifying individual thyroid nodules, which are inefficient and often do not take into consideration the situation when different neoplastic thyroid diseases co-exist. Besides, in the presence of a malignancy, the decision for subsequent treatment is relatively straightforward; usually, a thyroidectomy is indicated. However, in the absence of a malignancy, the most appropriate course of treatment can only be determined if all the pathology that may co-exist in the thyroid are known. In this context, here we introduce a multi-channel CNN architecture that can not only detects thyroid diseases for different lobes but also enhance the classification accuracy.

Performance comparison experiments of the four types of kernel size combinations showed that the combination of the kernel sizes of $1 \times 1$ and $7 \times 7$ led to the best performing and most stable model. In addition to the accuracy score, such models also achieve the highest precision, specificity, NPV, and F1 scores among the four models. Although there was a slight drop in the sensitivity for the multi-channel architecture, this is expected. This is because as the model's complexity increases, the sensitivity will be affected since the model is more likely to be over-fitting. Since the sensitivity of the multi-channel architecture was merely a bit lower than the single-channel architecture, it is acceptable. On the other hand, the results also highlight the impact of kernel size choice on the CNN model performance and suggest that combined use of a smaller and a larger kernel sizes could further enhance the CNN performance. Specifically, in terms of the F1 scores, the multi-channel architectures displayed higher scores than the single-channel architectures. Also, the single-channel CNN reached the highest accuracy score of 0.902 when $7 \times 7$ convolutional kernel operations were applied, whereas the multi-channel architecture ($1 \times 1$ and $7 \times 7$) further increased the CNN model performance (i.e., the accuracy of 0.909).

Additionally, the sensitivity scores appeared to be relatively lower among the six performance metrics for all the experiments. The primary class affecting the CNN sensitivity score is "thyroiditis", in which when this disease exists, the sensitivity score tends to be dragged down. Such a result is expected because thyroiditis can co-exist with other kinds of neoplastic thyroid diseases on the same side of the lobe and can confound the analysis. Considering that, during the training stage, labels were given to each image based on the dominant class, which might exist cases where thyroiditis is manifested on the image but not labeled as the class, thereby resulting in the lower sensitivity of the model.

Regarding the precision measure, the class "goiter" tends to exhibit the highest score than the other classes for both multi-channel and single-channel architectures. The reason behind this might be due to the multi-nodular goiter manifestation on CT images that are relatively distinct from the other types of nodules, leading to higher precision scores. Besides, when goiter and cancer co-exist, the single-channel architecture achieved a high precision score of 0.98, compared to that of 0.91 for the multi-channel architecture. When goiter and adenoma co-exist, the single-channel model attained a precision of 0.95, whereas the multi-channel reached a precision of 1. Therefore, the precision scores are quite optimistic when the class "goiter" exists.

The proposed multi-channel CNN produced enhanced diagnostic performance compared to the ordinary single-channel models. Another essential advantage of the multi-channel architecture is that it can also deal with situations where different types of thyroid disease co-exist. Despite the promising diagnostic performance, this study has certain limitations, many of which are related to labeling of the images. In particular, as medical data are complex, each CT image may have more than one label; this would make the labeling process troublesome and might lead to possible classification error rates. In the proposed study, the dominating class of the image was given as labels; it would be of interest to develop a new multi-class classification model to process multi-labeled images and provide further insight into image-based thyroid disease diagnosis. In addition, although CT images are able to provide some details of the thyroid other than ultrasonography, it is generally not the preferred modality to characterize thyroid diseases in the clinical setting. As a result, clinicians may find it difficult to accept a thyroid CAD system that is based on CT images. Therefore, future studies may incorporate other medical imaging modalities for the design of CAD systems.

## 8. Conclusion

In conclusion, this study proposed a new deep learning-based application in automating thyroid disease detection with increased diagnostic accuracy and efficiency than the subjective clinical diagnosis. We proposed a multi-channel CNN architecture to detect neoplastic thyroid disease, and also take into consideration situations where diseases co-exist. The proposed model has been evaluated and compared to the conventional single-channel CNN models, and the former one established better performance. In addition, the multi-channel architecture was also evaluated with respect to different gender groups, and the results demonstrate the promising generalizability of the proposed model.

We envision that more studies will be applied to rigorously examine the generalization and performance of such multi-channel CNN architecture for diagnosis of different types of diseases other than thyroid diseases in clinical settings. Together, with the increasing evidence of the feasibility of deep learning-based approaches, clinicians may be more confident and comfortable working with the use of artificial intelligence-based diagnostic tools to reduce their workloads and mitigate the diagnostic bias or human false-positive rates.

## Conflict of interest

All authors declare that there is no conflict of interest regarding the publication of this paper.

## Acknowledgements

This research did not receive any specific grant from funding agencies in the public, commercial, or not-for-profit sectors.